# Coexistence of orbital degeneracy lifting and superconductivity in iron-based superconductors

H. Miao[1], L.-M. Wang[2], P. Richard[1], S.-F. Wu[1], J. Ma[1], T. Qian[1], L.-Y. Xing[1], X.-C. Wang[1], C.-Q. Jin[1], C.-P. Chou[4,2], Z. Wang[3], W. Ku[2], and H. Ding[1*]

[1]Beijing National Laboratory for Condensed Matter Physics, and Institute of Physics, Chinese Academy of Sciences, Beijing 100190, China

[2]Condensed Matter Physics and Materials Science Department, Brookhaven National Laboratory, Upton, New York 11973, USA

[3]Department of Physics, Boston College, Chestnut Hill, Massachusetts 02467, USA

[4]Beijing Computational Science Research Center, Beijing 100084, China

**In contrast to conventional superconducting (SC) materials, superconductivity in high-temperature superconductors (HTCs) usually emerges in the presence of other fluctuating orders with similar or higher energy scales[1-5], thus instigating debates over their relevance for the SC pairing mechanism. In iron-based superconductors (IBSCs), local orbital fluctuations have been proposed to be directly responsible for the structural phase transition[6,7] and closely related to the observed giant magnetic anisotropy and electronic nematicity[8-12]. However, whether superconductivity can emerge from, or even coexist with orbital fluctuations, remains unclear. Here we report the angle-resolved photoemission spectroscopy (ARPES) observation of the lifting of symmetry-protected band**

**degeneracy, and consequently the breakdown of local tetragonal symmetry in the SC state of Li(Fe$_{1-x}$Co$_x$)As. Supported by theoretical simulations, we analyse the doping and temperature dependences of this band-splitting and demonstrate an intimate connection between ferro-orbital correlations and superconductivity.**

In IBSCs, the orbital degree of freedom is believed to play an important role and can be closely related to the emergence of superconductivity[13,14]. Ferro-orbital (FO) order, which leads to unequal occupation of $d_{xz}/d_{yz}$ orbitals, has also been proposed as the origin of electronic nematicity[6,7]. However, probing FO fluctuations directly in the absence of structural and magnetic phase transitions has never been explored, and whether FO fluctuations coexist or compete with the SC order is still an open question. In the tetragonal phase without spin-orbit coupling (SOC), $d_{xz}/d_{yz}$ orbitals are degenerate at the Brillouin zone centre (Γ point), which is guaranteed by point-group symmetry. The formation of FO ordering would lift the degeneracy at Γ, resulting in a band gap $\Delta_{band}$ that can be monitored directly by ARPES.

In addition of having a natural non-polar cleaving surface preserving its bulk properties[15-17], Li(Fe$_{1-x}$Co$_x$)As has neither structural nor magnetic phase transitions in its whole phase diagram[18], enabling us to study fluctuations in absence of long-range order. In Fig. 1, we compare the electronic band dispersion of LiFeAs and LiFe$_{0.88}$Co$_{0.12}$As around the Γ point. Our polarization analysis confirms that the α and α' bands, which are mainly composed of $d_{xz}/d_{yz}$ orbitals, have even and odd symmetries, respectively[19,28,29]. The extracted band dispersion[20] in LiFe$_{0.88}$Co$_{0.12}$As ($T_c$ = 4 K) indicates that both the α and α' bands sink below $E_F$ and are exactly degenerate at the Γ point, as required by symmetry. In contrast, the α' band crosses $E_F$ at $k_F$ = 0.03 π/a in the parent compound LiFeAs ($T_c$ = 18 K), whereas the top of the α band lies about 12 meV below $E_F$, which means that the $d_{xz}/d_{yz}$ orbitals are split in LiFeAs without long-range magnetic and orbital orders. To precisely resolve the band splitting, we recorded very high-energy resolution ARPES intensity plots of LiFeAs and LiFe$_{0.88}$Co$_{0.12}$As, as shown in Fig. 1. From the high-resolution data, we evaluate the band gap to $\Delta_{band}$ ~ 14 meV in LiFeAs by extrapolating the top of the α' band using a parabolic fit, and we confirm the degeneracy of the $d_{xz}/d_{yz}$ bands in LiFe$_{0.88}$Co$_{0.12}$As. By zooming near $E_F$, we find that the α' band further splits into two branches, as shown in Figs. 1c and 1g. While one branch is the continuous extension of the high binding energy dispersion, the other one shows an inflection point at 14 meV binding energy. As discussed later, the observed fine structure is consistent with the calculated electronic structure in the orbital nematic phase on twinned samples[21], and supports that the observed band splitting is caused by FO fluctuations. As reported previously[22], we distinguish an electron band at the Γ point of electron-doped LiFe$_{0.88}$Co$_{0.12}$As, which is not clear in the synchrotron-based results most likely due to different $k_z$ positions. We suspect that this small electron band has a strong As $p_z$ orbital component and is similar to the one observed in (Tl,Rb)$_y$Fe$_{2-x}$Se$_2$[23].

To check whether the $d_{xz}/d_{yz}$ splitting is a general feature of the IBSCs, we performed similar experiments on various materials and summarized the results in Fig. 2. We first considered LiFe$_{0.94}$Co$_{0.06}$As ($T_c$ = 10 K), which has an intermediate doping between LiFeAs and LiFe$_{0.88}$Co$_{0.12}$As[29]. Unlike in LiFeAs and similarly to LiFe$_{0.88}$Co$_{0.12}$As, the α' band falls below $E_F$. For this material, we find $\Delta_{band}$ = 10 meV, suggesting that the band splitting is gradually supressed as $T_c$ decreases from 18 K to 4 K. We also studied the band splitting in NaFe$_{0.95}$Co$_{0.05}$As ($T_c$ = 18 K)[24], which is isostructural to LiFeAs (so-called 111 structure). From the extracted band dispersions displayed in Fig. 2d, we deduce that $\Delta_{band}$ = 15 meV in this particular compound. Interestingly, all our data on the 111 crystal structure indicate that $\Delta_{band}$ scales with $T_c$ in this family of materials, suggesting that the band splitting might be related to superconductivity. Interestingly, there is at least one other IBSC for which a $d_{xz}/d_{yz}$ band splitting is clearly observed. Indeed, this observation has been reported for the FeTe$_{1-x}$Se$_x$ family of IBSCs[25,26]. Using the data from Miao et al.[25], reproduced in Fig. 2e, we find that $\Delta_{band}$ = 18 meV in FeTe$_{0.55}$Se$_{0.45}$, which is even larger than in LiFeAs. The observed band splitting in all the IBSCs studied here strongly suggests that the $d_{xz}/d_{yz}$ separation at the $\Gamma$ point has a fundamental origin.

We now focus on the temperature evolution of $\Delta_{band}$ in LiFeAs. For this purpose, we show in Fig. 3 high energy resolution ARPES cuts across the $\Gamma$ point recorded between 50 and 250 K. The data are divided by the Fermi-Dirac function convoluted by the resolution function to reveal the band dispersion above $E_F$, which are obtained from parabolic fits. While the line width of the α and β bands broaden with temperature, their dispersions are unaffected. The α' band, on the other hand, gradually shifts downward and its top almost merges with that of the α band at 250 K. In Fig. 2m, we show the evolution of $\Delta_{band}$ as function of temperature by using different methods that all show that the $d_{xz}/d_{yz}$ splitting decreasing gradually from nearly 0 at 250 K to about 14 meV at 50 K[29]. Interestingly, the splitting survives even below the SC phase transition at 18 K. As shown in Figs. 2n - 2p, the α' band opens up a SC gap below $T_c$, whereas the α band is barely changed. This observation proves that the band splitting coexists with superconductivity.

The observation of the degeneracy lifting of $d_{xz}/d_{yz}$ orbitals in the tetragonal phase is puzzling. In order to demonstrate its origin, we investigate a simple model of quasi-1D electronic system (since $d_{xz}/d_{yz}$ orbitals have strongly anisotropic quasi-1D hopping integral) under the influence of a spatially fluctuating local FO order parameter (represented by a diagonal Ising field)[28,29], and we display the results in Fig. 4. When the local order parameter has only short-range correlations (exponential decay), no clear indication of the fluctuating order is observed in the electronic structure other than the scattering of the particle that broadens the spectral function, as illustrated in the top row of Fig. 4. In contrast, when the spatial correlations of the local order parameter are long-ranged (power law decay), the quasiparticle peak splits in two and a pseudogap in the spectral function develops in between, as shown in Figs. 4f – 4h. This pseudogap corresponds to the splitting of the degenerate bands shown in Fig. 1. Although the spectral function exhibits features identical to those expected in the presence of a macroscopic long-range order, we emphasize that the system has not yet developed a true order, but only long-range spatial correlations. In other words, the one-particle Green's function has gone ahead and reflects the underlying "almost ordered" electronic structure. Therefore, the experimentally observed doping-dependent splitting between the $d_{xz}/d_{yz}$ bands in the absence of FO order can be attributed to strong, slow-decaying, long-range FO correlations that cover a large region of the phase diagram and eventually support superconductivity at low temperature.

Very recently, an electronic Raman scattering study of Ba(Fe$_{1-x}$Co$_x$)$_2$As$_2$[27] and a combined study of magnetic-torque and X-ray in BaFe$_2$(As$_{1-x}$P$_x$)$_2$[10] reported electronic nematicity in the absence of a magnetic phase transition. The observed "nematic" signal persists far above the SC and structural phase transitions, indicating the presence of a strong fluctuating orbital order, which is consistent with our doping- and temperature-dependent results. In the present study, we find that superconductivity emerges in the regime of strong FO correlations and that the SC transition temperature scales with the strength of the local orbital order, suggesting an intimate connection between FO correlations and superconductivity. This observation is very important since a recent ARPES and NMR study reports enhancement of low-energy antiferromagnetic spin fluctuations in LiFe$_{1-x}$Co$_x$As samples with lower $T_c$[22], thus suggesting that such low-energy antiferromagnetic fluctuations alone do not control the strength of superconductivity in LiFe$_{1-x}$Co$_x$As.

Finally, we discuss the effect of spin-orbit coupling (SOC). In principle, SOC can lift the degeneracy of the $d_{xz}/d_{yz}$ orbitals at the $\Gamma$ point while maintaining local tetragonal symmetry. However, since SOC is a local effect and barely changes with doping and temperature, our observation of $\Delta_{band}$ variations as a function of doping and temperature is inconsistent with this scenario. Moreover, the clearly resolved inflection point on one branch of the fine structure shown in Fig. 1c is consistent with the calculated electronic structure in the orbital nematic phase on twinned samples[21], and therefore supports our assumption that the observed $d_{xz}/d_{yz}$ splitting is caused by FO fluctuations instead of SOC. Although spin fluctuations have been widely studied and discussed, fewer experimental studies on the orbital fluctuations can be found in the literature. Our study provides evidence of strong, long-range FO correlations in

IBSCs and demonstrates their intimate connection with superconductivity in LiFe$_{1-x}$Co$_x$As.

**Methods:**

Single crystals of LiFe$_{1-x}$Co$_x$As were synthesized by a self-flux method using Li$_3$As, Fe$_{1-x}$Co$_x$As and As powders as the starting materials. The Li$_3$As, Fe$_{1-x}$Co$_x$As and As powders were weighed according to the element ratio of Li(Fe$_{1-x}$Co$_x$)$_{0.3}$As. The mixture was grounded and put into alumina crucible and sealed in Nb crucibles under 1 atm of Argon gas. The Nb crucible was then sealed in an evacuated quartz tube, heated to 1100 °C and slowly cooled down to 700 °C at a rate of 3 °C/hr. High energy resolution ARPES data were recorded at the Institute of Physics, Chinese Academy of Sciences, using the He I$\alpha$ ($h\nu$ = 21.2 eV) resonance line of an helium discharge lamp. The angular and momentum resolutions were set to 0.2° and 3 meV, respectively. ARPES polarization measurements were performed at beamlines PGM and Apple-PGM of the Synchrotron Radiation Center (Wisconsin) equipped with a Scienta R4000 analyzer and a Scienta SES 200 analyzer, respectively. The energy and angular resolutions were set at 20 meV and 0.2°, respectively. All samples were cleaved *in situ* and measured in a vacuum better than 3*10$^{-11}$ Torr.

**Acknowledgement:**

We thank A. Tsvelik, J.-P. Hu, F. Wang, P. D. Johnson and H.-B. Yang for useful discussions. This work was supported by grants from CAS (NO. 2010Y1JB6), MOST (Nos. 2010CB923000, 2011CBA001000, 2013CB921703), NSFC (11004232 and 11274362). Theoretical study is supported

by US Department of Energy, Office of Science DE-AC02-98CH10886. This work is based in part on research conducted at the Synchrotron Radiation Center, which is primarily funded by the University of Wisconsin-Madison with the University of Wisconsin-Milwaukee.

**Author contributions:**

H. M., T.Q., S.-F. W., and J. M. carried out the experiments; H. M. analyzed the data; L. -Y. X., X.-C. W. and C.-Q. J. provided samples; L.-M. W., C.-P. C., A.T and W. K. carried out the theoretical calculations; H. M., P. R., T. Q., L.-M. W., W. K. and H. D. wrote the paper. All authors discussed the results and commented on the manuscript.

**Additional information:**

The authors declare no competing financial interests.

**Figure 1 | Observation of split $d_{xz}/d_{yz}$ bands in LiFeAs and degenerate $d_{xz}/d_{yz}$ bands in LiFe$_{0.88}$Co$_{0.12}$As. a** and **e,** ARPES intensity plots of LiFeAs measured along the Γ-M direction, and recorded with 51 eV incident light in the σ and π configurations to select odd and even orbital symmetries[19,28,29], respectively. **b** and **f,** Same as panels **a** and **e**, but for LiFe$_{0.88}$Co$_{0.12}$As. **c** and **d,** High energy resolution ARPES cuts along the Γ-M direction of LiFeAs and LiFe$_{0.88}$Co$_{0.12}$As, respectively, recorded with the He Iα line of an helium discharge lamp. The red and blue curves in panel **a**, **b**, **e** and **f** are the original and fitted momentum distribution curves at $E_F$, respectively. The red, blue and green circles represent the peak positions associated with the α ($d_{even}$), α' ($d_{odd}$) and β ($d_{xy}$) bands, respectively, where $d_{odd}(d_{even})$ is the odd(even) linear combination of the $d_{xz}$ and $d_{yz}$ orbitals. By comparing the extracted band dispersions, we conclude that the band top of the $d_{xz}/d_{yz}$ bands are split in LiFeAs and degenerate in LiFe$_{0.88}$Co$_{0.12}$As. The zoom on the high energy resolution data of **c** reveals that the α' band is further split into two branches, as shown in the inset. **g** and **h**, EDCs corresponding to the data shown in panels **c** and **d**, respectively. The blue EDC in panel **g**, also shown in inset, illustrates the splitting of the α' band.

**Figure 2 | Doping and material dependence of the $d_{xz}/d_{yz}$ band splitting. a – c,** Extracted band dispersion of the $d_{xz}/d_{yz}$ bands in LiFeAs, LiFe$_{0.94}$Co$_{0.06}$As[29], and LiFe$_{0.88}$Co$_{0.12}$As, respectively. **d** and **e**, Extracted band dispersion of NaFe$_{0.95}$Co$_{0.05}$As[24] and FeTe$_{0.55}$Se$_{0.45}$[25], respectively. Red dashed curves are parabolic fits. **f**, Doping and $T_c$ dependence of $\Delta_{band}$. The open and plain symbols refer to the doping (bottom) and $T_c$ (top) axes. Error bars are determined by standard deviation of the fitting parameters[29].

**Figure 3 | Temperature evolution of the band gap $\Delta_{band}$ and its coexistence with superconductivity. a – e,** High energy resolution ARPES intensity plots at T = 250 K, 200 K, 150 K, 100 K and 50 K, respectively. **f – j,** Same data but divided by the Fermi-Dirac function convoluted with the system resolution function to probe the electronic structure above $E_F$. Band dispersions at different temperatures are extracted using momentum distribution curves (MDCs) and fitted to parabolic functions. **k,** Energy distribution curves (EDCs) at the Γ point at different temperatures. We used two Lorentzian peaks to extract the top of the α and α' bands, and plot the fitted results on top of the original data using red dashed curves[29]. The top of the α' band is shifted towards high binding energy as temperature increases, and it almost merges with the top of the α band at 250 K. **l,** MDCs recorded 20 meV below $E_F$, which corresponds to the red dashed lines shown in **a – e**. At high temperature, the α' peak positions move towards Γ, indicating that the band moves downward in energy. In contrast, the peak positions of the α and β bands are unchanged[29]. **m,** temperature evolution of $\Delta_{band}$. The values of $\Delta_{band}$ are extracted from the electronic band dispersions, EDCs and MDCs. Error bars are determined by the standard deviation of the fitted parameters. **n** and **o** are ARPES intensity plots just above and well below $T_c$, respectively. **p,** EDCs at the Γ point above and below $T_c$.

**Figure 4 | Sensing long-range correlations without real order.** In a typical mean-field treatment, the effects of the order parameter, for example a gap opening, are only visible in the presence of long-range order, represented by a non-zero order parameter. Without a real long-range order, the local order parameter only scatters the carriers and smears the spectral function. This is illustrated by the upper panels: **a,** an example of the configuration of a FO disordered system (represented by an Ising field) without long-range correlations, **b,** the resulting average band structure that couples to the Ising field, showing no clear splitting of the band. **c**, **d**, the average spectral function at momentum **k**=(0,0) and **k**=(0.25,0.25). However, the situation becomes interesting when the disordered system contains long-range, power-law decaying correlations of the local order parameter. The lower panels **e-h** illustrates this interesting FO disordered case. One finds that even though the order parameter remains zero, the average spectral function now demonstrates clear signatures of coupling to the local order. For example, a splitting of degeneracy is observed in **f**, and a pseudo-gap is found in **g**, both of which being expected in the presence of an order parameter, but now already occur in the disordered system with long-range correlations.

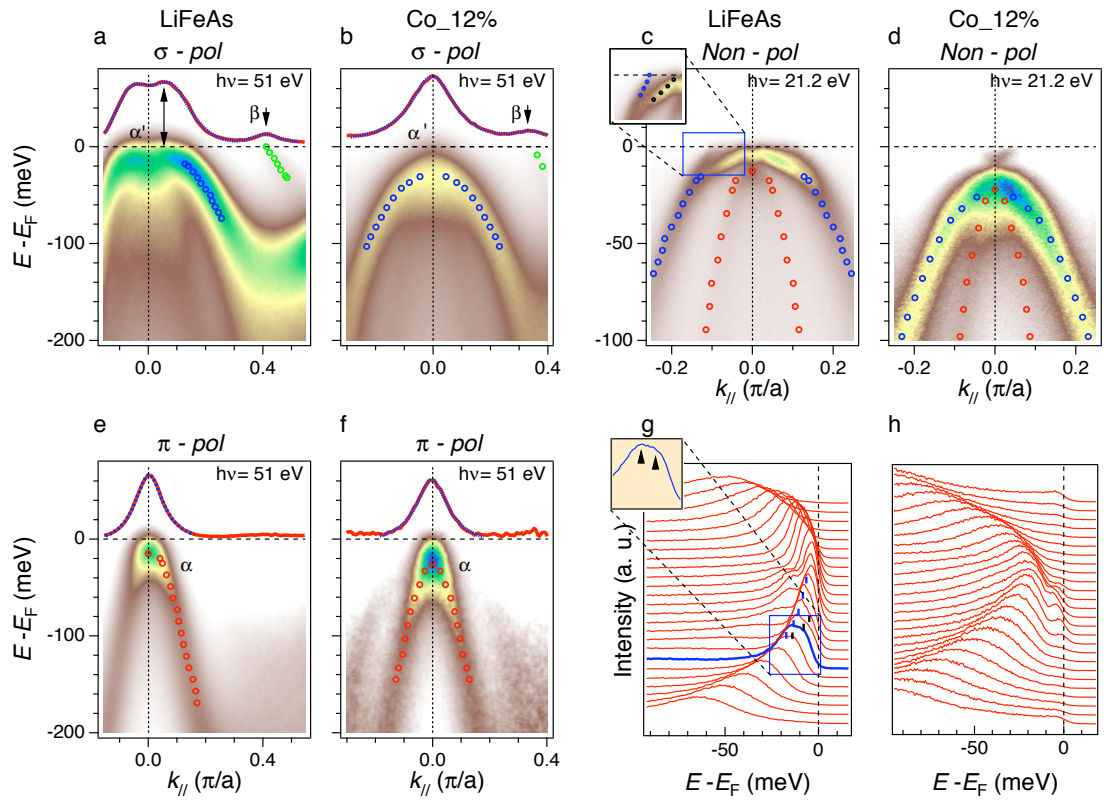

Figure 1

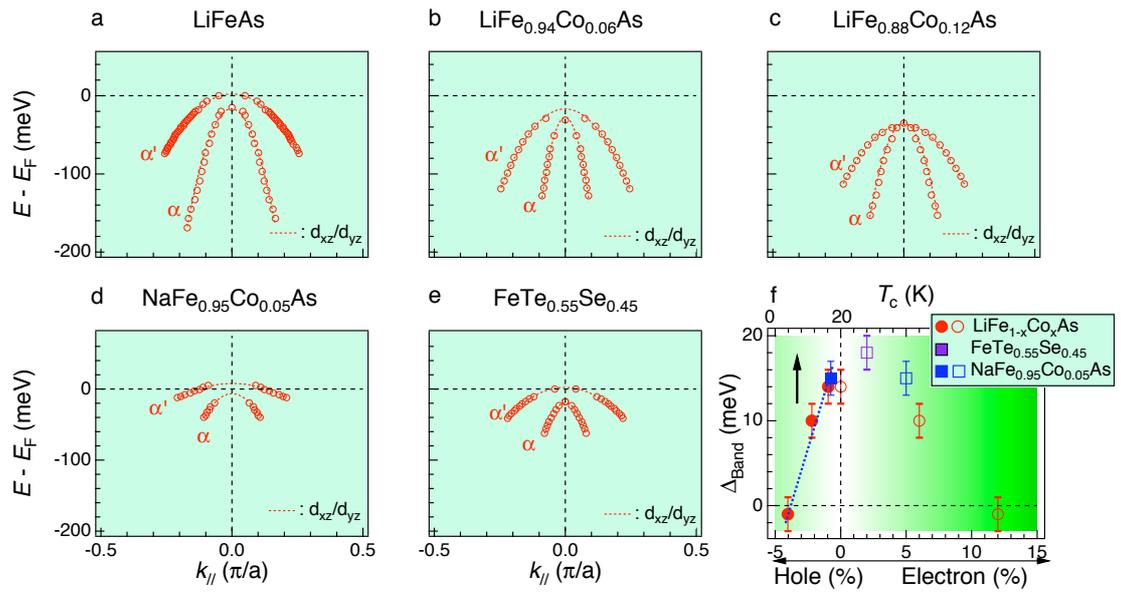

Figure 2

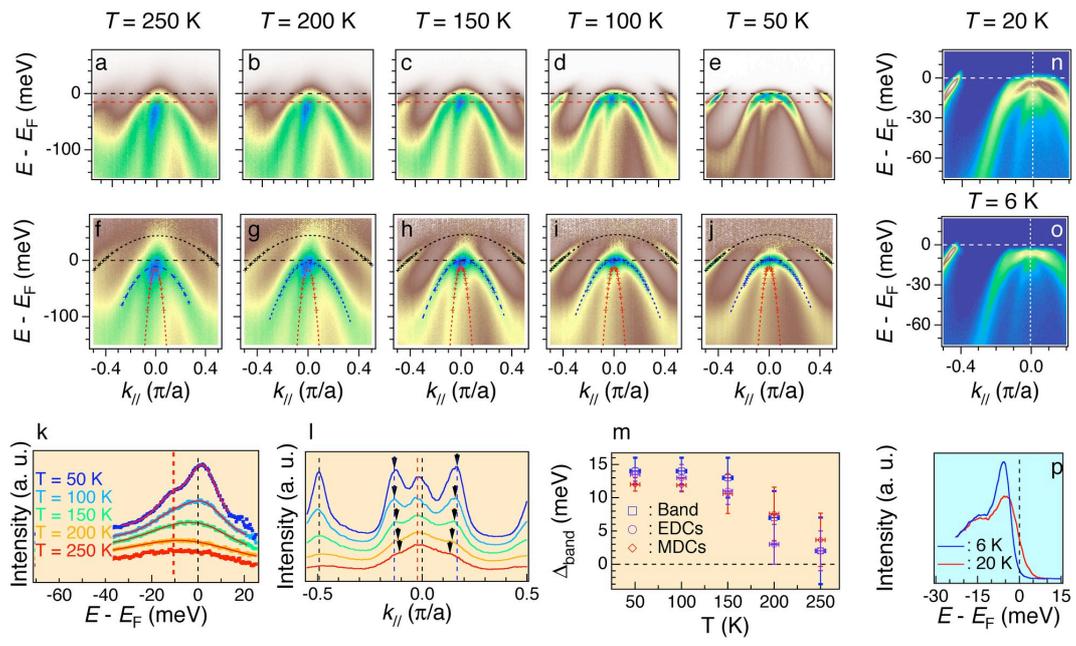

Figure 3

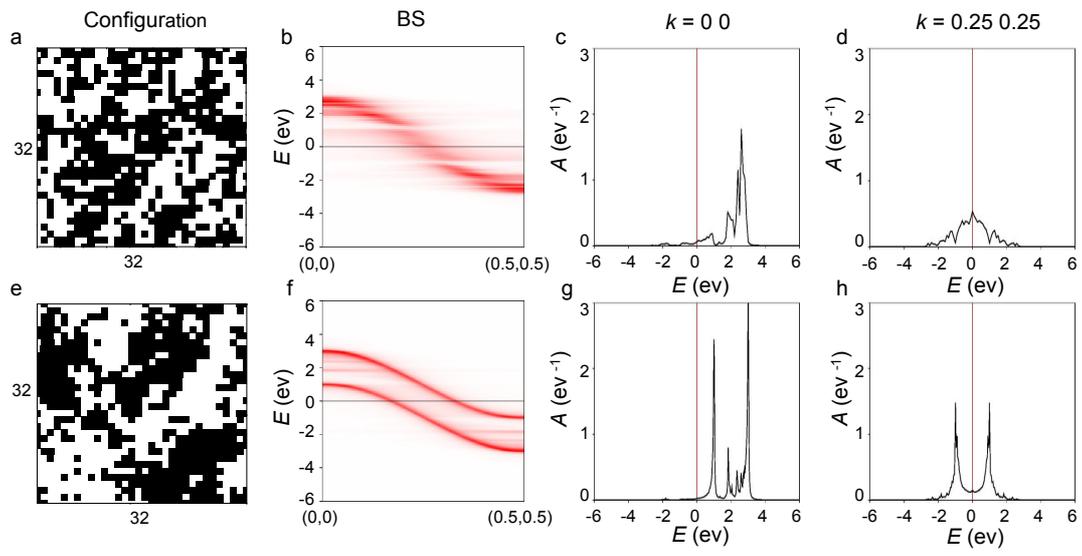

Figure 4